\documentclass[journal=jacsat,manuscript=article]{achemso}

\usepackage[version=3]{mhchem} 
\usepackage[T1]{fontenc}       

\author{Jinwoong Hwang}
\affiliation{Department of Physics, Pusan National University, Busan 46241, South Korea}
\alsoaffiliation{Advanced Light Source, Lawrence Berkeley National Laboratory, Berkeley, California 94720, USA}

\author{Seungseok Lee}
\affiliation{Center for Complex Phase Materials, Max Planck POSTECH/Korea Research Initiative, Pohang 37673, South Korea}
\alsoaffiliation{Division of Advanced Material Science, Pohang University of Science and Technology, Pohang 37673, South Korea}

\author{Ji-Eun Lee}
\affiliation{Department of Physics, Pusan National University, Busan 46241, South Korea}
\alsoaffiliation{Center for Spintronics, Korea Institute of Science and Technology, Seoul 02792, South Korea}

\author{Minhee Kang}
\affiliation{Department of Physics, Pusan National University, Busan 46241, South Korea}

\author{Hyejin Ryu}
\affiliation{Advanced Light Source, Lawrence Berkeley National Laboratory, Berkeley, California 94720, USA}
\alsoaffiliation{Center for Complex Phase Materials, Max Planck POSTECH/Korea Research Initiative, Pohang 37673, South Korea}
\alsoaffiliation{Center for Spintronics, Korea Institute of Science and Technology, Seoul 02792, South Korea}

\author{Hyun-Jeong Joo}
\affiliation{Department of Physics, Pusan National University, Busan 46241, South Korea}

\author{Jonathan Denlinger}
\affiliation{Advanced Light Source, Lawrence Berkeley National Laboratory, Berkeley, California 94720, USA}

\author{Jae-Hoon Park}
\affiliation{Center for Complex Phase Materials, Max Planck POSTECH/Korea Research Initiative, Pohang 37673, South Korea}
\alsoaffiliation{Division of Advanced Material Science, Pohang University of Science and Technology, Pohang 37673, South Korea}
\alsoaffiliation{Department of Physics, Pohang University of Science and Technology, Pohang 37673, South Korea}
\email{jhp@postech.ac.kr}

\author{Choongyu Hwang}
\affiliation{Department of Physics, Pusan National University, Busan 46241, South Korea}
\email{ckhwang@pusan.ac.kr}

\title
  {Tunable Kondo resonance at a pristine two-dimensional Dirac semimetal on a Kondo insulator}

\abbreviations{IR,NMR,UV}

\begin{document}

%
%
%
%
%
%
\begin{abstract}

Proximity of two different materials leads to an intricate coupling of quasiparticles so that an unprecedented electronic state is often realized at the interface. Here, we demonstrate a resonance-type many-body ground state in graphene, a non-magnetic two-dimensional Dirac semimetal, when grown on SmB$_6$, a Kondo insulator, via thermal decomposition of fullerene molecules. This ground state is typically observed in three-dimensional magnetic materials with correlated electrons. Above the characteristic Kondo temperature  of the substrate, the electron band structure of pristine graphene remains almost intact. As temperature decreases, however, the Dirac fermions of graphene become hybridized with the Sm 4$f$ states. Remarkable enhancement of the hybridization and Kondo resonance is observed with further cooling and increasing charge carrier density of graphene, evidencing the Kondo screening of the Sm 4$f$ local magnetic moment by the conduction electrons of graphene at the interface. These findings manifest the realization of the Kondo effect in graphene by the proximity of SmB$_6$ that is tuned by temperature and charge carrier density of graphene.

\end{abstract}

\section{Introduction}

Formation of an interface is one of the important ingredients in realizing emergent phenomena and manipulating device functionality. For this purpose, an atomically thin two-dimensional (2D) crystal is an ideal platform when forming an interface with a three-dimensional (3D) exotic substrate, since its property is easily affected by an environment~\cite{Bae}. Indeed, various intriguing phenomena have been reported in the studies of 2D crystals placed on 3D substrates, for example, the proximity-induced superconductivity~\cite{Heersche,Kessler} and magnetism~\cite{Zhong}, and the enhancement of spin conductivity~\cite{Shi}, superconductivity~\cite{Wang,Ge}, electron correlations~\cite{RyuSTO}, and spin-orbit coupling~\cite{Dedkov,Gmitra,Frank}. In addition, graphene on magnetic materials exhibits characteristic spin properties such as single-spin Dirac cone~\cite{Usachov}, Hedgehog spin structure~\cite{Varykhalov}, and large exchange splitting~\cite{Wu}. Hence, interfacing 2D system is a powerful methodology not only to explore exotic many-body phenomena confined in a single atomic layer limit.

One of the renowned many-body phenomena is the Kondo effect~\cite{Kondo}. The Kondo effect refers to the phenomena in which a resonance-type hybridized many-body ground is induced in a non-magnetic metallic system in the presence of a local magnetic moment. The interaction between the local magnetic moment and the spins of surrounding electrons from the metallic host leads to antiferromagnetic spin alignment to screen the local magnetic moment, resulting in the formation of the Kondo resonance state. Since its first discovery, the Kondo effect has been one of the key elements to understand many-body effects in a solid system especially related with its magnetic properties. The Kondo effect has been well studied in 3D materials such as dilute magnetic alloys~\cite{Kondo_book} and heavy fermionic compounds~\cite{Coleman}, while it has been also evidenced in lower dimensions, including the surface of 3D metals~\cite{Knorr}, carbon nanotubes~\cite{Nygard}, quantum dots~\cite{Cronenwett}, and single-electron transistors~\cite{Gorden}. However, despite the advance in various wan der Waals materials, the Kondo effect in the prototypical 2D crystals has been barely paid attention, hindering to understand the nature of 2D Kondo physics~\cite{Zitko,Defect,Ripple,Barua,Fritz2}.  In particular, the low-dimensional systems have been utilized to investigate an impurity-type Kondo effect, i.\,e.\,, Kondo effect induced by a single magnetic impurity, leaving a lattice-type low-dimensional Kondo effect unveiled.

The {\it in-situ} thermal decomposition of fullerene molecules on a 3D Kondo insulator can be an ideal approach to realize and investigate 2D lattice-type Kondo effect. This approach makes it possible not only to prepare an interface between graphene and a 3D substrate of interests, as demonstrated in spectroscopic~\cite{Cepek,Perdigao,Azpeitia} and microscopic~\cite{Lu,Yamada,Fei} studies for transition metals, but also to align crystalline orientation of the two ingredients of the interface that cannot be achieved by the well-known method of preparing a 2D/3D interface, e.\,g.\,, mechanical transfer. The latter is crucial in device applications, because the characteristics of the interaction between a 2D crystal and a 3D substrate varies significantly in realistic devices, depending on crystalline orientation at the interface~\cite{Zhou}. Furthermore, the mechanical transfer process deteriorates the surface of a substrate due to contamination inevitable during the transfer process, which blows up the intrinsic properties of the substrate at its interface~\cite{RyuSTO}. 

Here we demonstrate that the formation of an interface between two exotic materials, a 2D Dirac semimetal, graphene, and a 3D Kondo insulator, SmB$_6$. Low-energy-electron diffraction (LEED) and angle-resolved photoemission spectroscopy (ARPES) studies show a well-defined interface between graphene and SmB$_6$ via the thermal decomposition method with their crystalline orientation perfectly aligned to each other. Surprisingly, at the interface, not only the spectral intensity of the Kondo resonance, stemming from the substrate, but also the kink induced by the hybridization between the Kondo resonance and the metallic band of overlying graphene exhibit characteristic dependence on both temperature and charge carrier density of graphene, which exactly follow the expectation from Kondo physics. These findings provide an experimental evidence that the quasiparticles from two exotic materials exhibit magnetic coupling to form a new many-body ground state at their interface, which is tuned by temperature and charge carrier density.

\section{Results and discussion}

Figure~1(a) shows the schematics of the growth of graphene on SmB$_6$ using the thermal decomposition method to investigate the interplay between a 2D Dirac semimetal~\cite{Neto} and a 3D Kondo insulator~\cite{Jonathan1,Jonathan2,Jiang}. The left panel of Fig.~1(b) shows a rectangular 1$\times$1 LEED pattern, corresponding to the clean SmB$_6$(110) surface as denoted by blue circles. After the graphene growth, additional hexagonal spots corresponding to graphene appear in the LEED pattern as denoted by red circles in the right panel of Fig.~1(b). Although the rectangular pattern of SmB$_6$ is not commensurate with the hexagonal one of graphene, their azimuthal orientations are aligned with each other. 

  \begin{figure*}[t]
  \begin{center}
  \includegraphics[width=1\columnwidth]{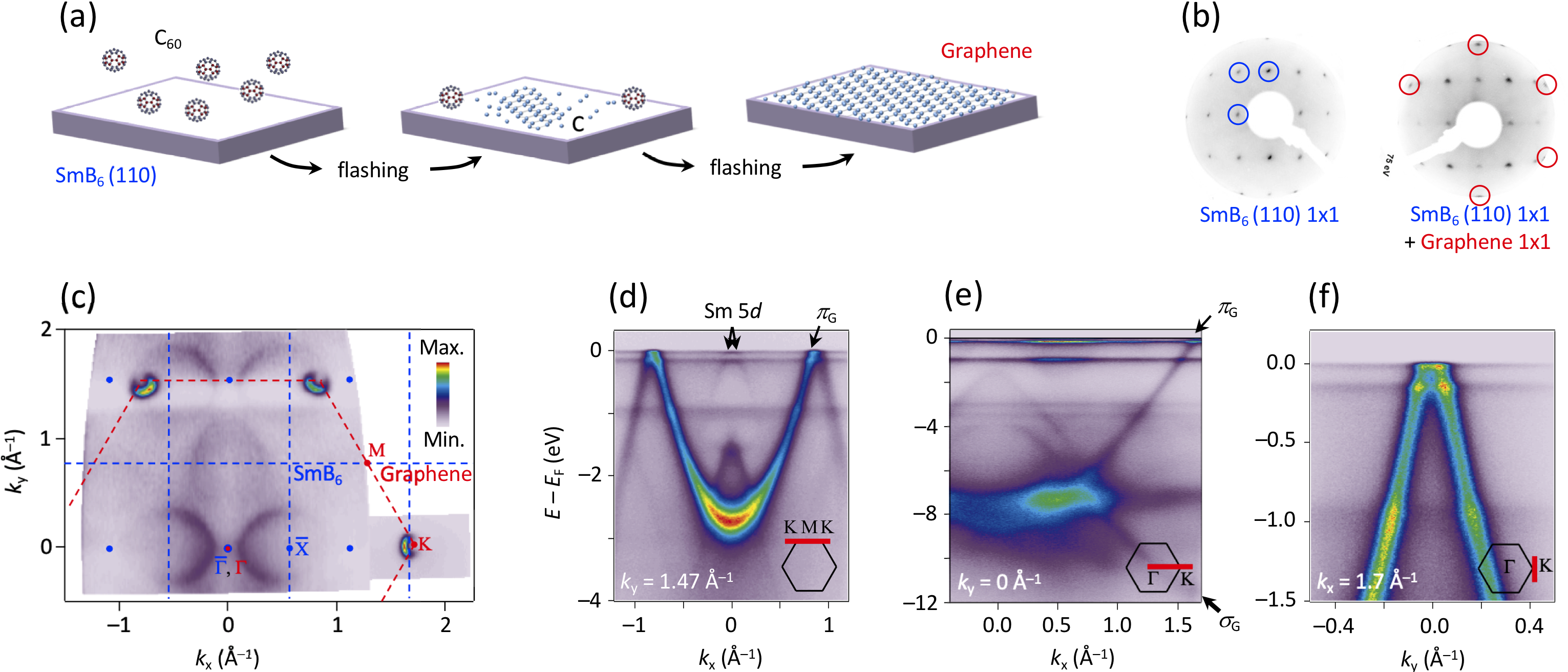}
  \end{center}
  \caption{(a) Schematics of the thermal decomposition process of fullerene molecules on SmB$_6$ to form graphene. (b) LEED patterns of SmB$_6$ (left panel) and graphene on SmB$_6$ (right panel). (c) A constant energy ARPES intensity map of graphene on SmB$_6$ taken at 0.5~eV below $E_{\rm F}$ at 27~K. Blue-dashed rectangle and red-dashed hexagon are the unit cell of SmB$_6$ and graphene, respectively. (d-e) ARPES intensity maps of graphene on SmB$_6$ measured at 27~K along the $k_{\rm x}$-axis at $k_{\rm y}$=1.47~\AA$^{-1}$ and $k_{\rm y}$=0. $\pi_{\rm G}$ and $\sigma_{\rm G}$ denote the $\pi$ and $\sigma$ bands of graphene, respectively. (f) An ARPES intensity map taken at 41~K along the $k_{\rm y}$-axis at $k_{\rm x}$=1.7~\AA$^{-1}$.}
  \label{Fig1}
  \end{figure*}

The hexagonal pattern of the add-on graphene is also observed in the electron band structure measured by ARPES. Figure~1(c) shows a constant energy ARPES intensity map at 0.5~eV below the Fermi energy, $E_{\rm F}$. The blue-dashed lines are the Brillouin zone boundary of SmB$_6$(110) that crosses its characteristic oval-shaped electron pockets~\cite{Jonathan_110}. The constant energy map exhibits an additional hexagonal feature from graphene denoted by red-dashed lines, which are not relevant to the high-symmetry points of SmB$_6$. 
This indicates that overlying graphene is not commensurate with the substrate, whereas their azimuthal orientations are aligned with each other in agreement with the LEED result (Fig.~1(b)). When the graphene/SmB$_6$ interface is a vertical heterostructure as depicted in the cartoon in Fig.~1(a), the ARPES and LEED data manifest themselves that in-plane orientation of graphene exactly follows that of SmB$_6$. The crescent-like shape observed at the constant energy map originates from the Berry phase of quasiparticles in graphene~\cite{CK}. ARPES intensity maps along the $k_{\rm x}$-axis at $k_{\rm y}=1.47$~\AA$^{-1}$ (Fig.~1(d)) and $k_{\rm y}=0$ (Fig.~1(e)) also show the electron band structure of graphene ($\pi_{\rm G}$ and $\sigma_{\rm G}$) along the K-M-K and $\Gamma$-K directions of the graphene unit cell as denoted in insets, respectively. While graphene exhibits the characteristic conical dispersion, a close look at the band structure around $E_{\rm F}$ does not show the atop of the conical shape or the crossing point of the $\pi$ band, so-called Dirac energy, $E_{\rm D}$, (Fig.~1(f)), indicating that graphene is charge doped by a small amount of holes ($n_{\rm h}=2.0\times10^{11}~{\rm cm}^{-2}$) with respect to its natural form in which $E_{\rm D}$ is located at $E_{\rm F}$. The observation of the electron band structure of both SmB$_6$(110) and graphene indicates that a clean interfacial structure is successfully formed without ruining their intrinsic properties.

SmB$_6$ is a well-known Kondo insulator, where the Sm 4$f$ local magnetic moment is screened by the antiferromagnetic spin alignment of delocalized Sm 5$d$ electrons~\cite{Nickerson,Patthey,Cooley}. Such spin alignment leads to the formation of a Kondo resonance state. However, at high temperature, due to the competition between the thermal and quantum spin-flipping fluctuations, the spins of the conduction electrons of the metallic host are not able to screen the local magnetic moment and hence the metallic band remains intact as schematically shown in Fig.~2(a). Upon decreasing temperature, the spin-flipping scattering becomes prominent compared to phonon scattering, so that some of the metallic electrons participate in the screening of the local magnetic moment, leading to the formation of an incoherent Kondo resonance state as shown in Fig.~2(b). At very low temperature, especially below a characteristic temperature called Kondo temperature $T_{\rm K}$, the local magnetic moment is fully screened by the spins of the conduction electrons, opening a hybridization gap close to $E_{\rm F}$ as shown in Fig.~2(c). 

  \begin{figure*}[t]
  \begin{center}
  \includegraphics[width=1\columnwidth]{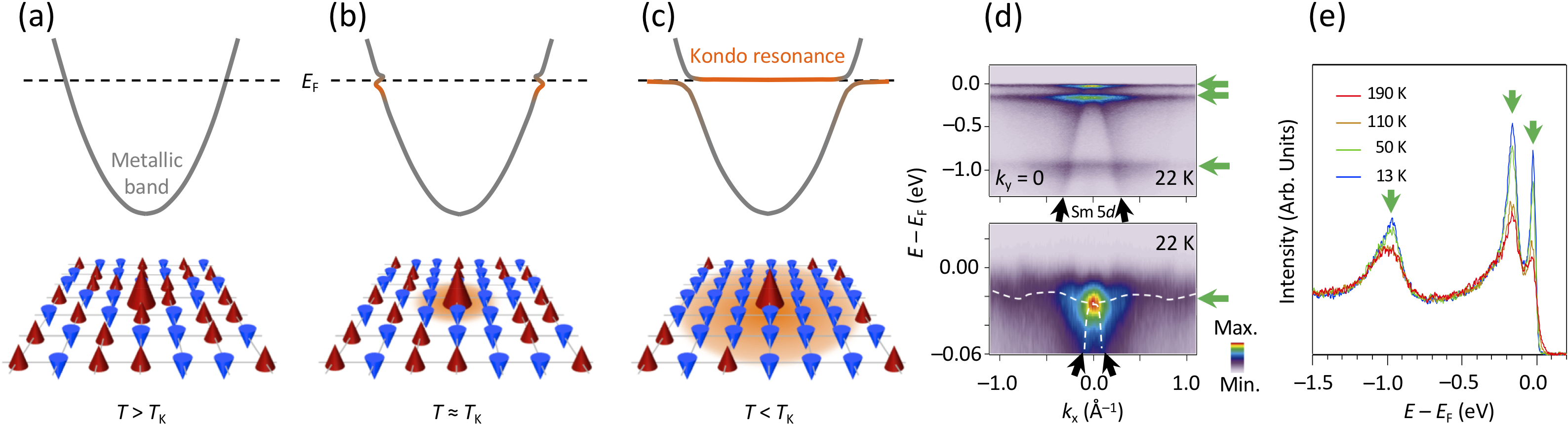}
  \end{center}
  \caption{(a-c) Cartoons of the evolution of a hybridized many-body ground state, Kondo resonance, in terms of the electron band structure and the spin configuration of a local magnetic moment (larger red cone) and conduction electrons (smaller blue and red cones). (d) An ARPES intensity map taken at 22~K along the $k_{\rm x}$-axis at $k_{\rm y}$=0. The lower panel is a zoomed-in view of the ARPES intensity map close to $E_{\rm F}$. The hybridized states between the lowest energy Kondo resonance states and the Sm 5$d$ band are denoted by white-dashed lines. The green and black arrows denote Kondo resonance states and the Sm 5$d$ band, respectively. (e) Temperature-dependent angle-integrated spectral intensity of the Kondo resonance states. }
  \label{Fig2}
  \end{figure*}

The drastically enhanced resistance of SmB$_6$ below $T_{\rm K}\approx150~K$ originates from this hybridization gap~\cite{Nickerson,Patthey}. Indeed, at 22~K, well below $T_{\rm K}$, the ARPES intensity map in the upper panel of Fig.~2(d) shows sharp Kondo resonance states at $E-E_{\rm F}=-0.02$ and $-0.17$~eV, corresponding to so-called $^{6}H_{5/2}$ and $^{6}H_{7/2}$ Sm 4$f$ final-state multiplets, respectively, and at $E-E_{\rm F}=-0.96$~eV, corresponding to so-called $^{6}F_{5/2}$ and $^{6}F_{7/2}$ Sm 4$f$ final-state multiplets, as denoted by green arrows~\cite{Jonathan2}. These Kondo resonance states hybridize with the Sm 5$d$ band, denoted by black arrows. Especially as shown in a zoomed-in view close to $E_{\rm F}$ (the lower panel of Fig.~2(d)), the Sm 5$d$ band does not cross $E_{\rm F}$ due to hybridization with the lowest energy Kondo resonance state, i.\,e.\,, the $^{6}H_{5/2}$ multiplet (hybridized states are denoted by dashed lines), manifesting the insulating behavior of SmB$_6$ at low temperature. Figure~2(e) displays angle-integrated spectral intensities of the Kondo resonance states at different temperatures. The intensity gradually increases as temperature decreases below $T_{\rm K}$, demonstrating the characteristic feature of the Kondo resonance~\cite{Jonathan2,Nickerson}.

The Kondo resonance states of SmB$_6$ exhibit dependence on charge carrier density of graphene. Since $T_{\rm K}$ is expected to be proportional to ${\rm exp}(-1/J\rho_{\rm F})$, where $J$ is the magnetic coupling and $\rho_{\rm F}$ is the density of states at $E_{\rm F}$, a change in the carrier will result in different $T_{\rm K}$ and hence spectral weight of the Kondo resonance.
Figure~3(a) shows an ARPES intensity map of graphene on SmB$_6$ taken at 24~K along the K-M-K directions of the graphene unit cell. When potassium is deposited on the sample, the graphene $\pi$ band gradually shifts towards the higher binding energy side, while the overall electron band structure of SmB$_6$ remains at almost the same binding energy, as shown in Figs.~3(b-d). This selective shift indicates that potassium dose only changes the charge carrier density of graphene, while that of SmB$_6$ remains almost the same. Surprisingly, the spectral weight of the Kondo resonance states denoted by green arrows becomes heavier with increasing charge carrier density of graphene, despite they originate from the substrate whose charge carrier density remains almost the same. To see the increase qualitatively, Fig.~3(e) shows the spectral intensity of the area denoted by a dashed rectangle in Fig.~3(a) as a function of potassium dose. The Kondo resonance state, especially the $^{6}H_{7/2}$ final-state multiplet, whose intensity is presented in the lower panel, increases in spectral intensity with increasing charge carrier density of graphene. As a result, the Kondo resonance states, despite stemming from SmB$_6$, are coupled to overlying graphene.

  \begin{figure*}
  \begin{center}
  \includegraphics[width=1\columnwidth]{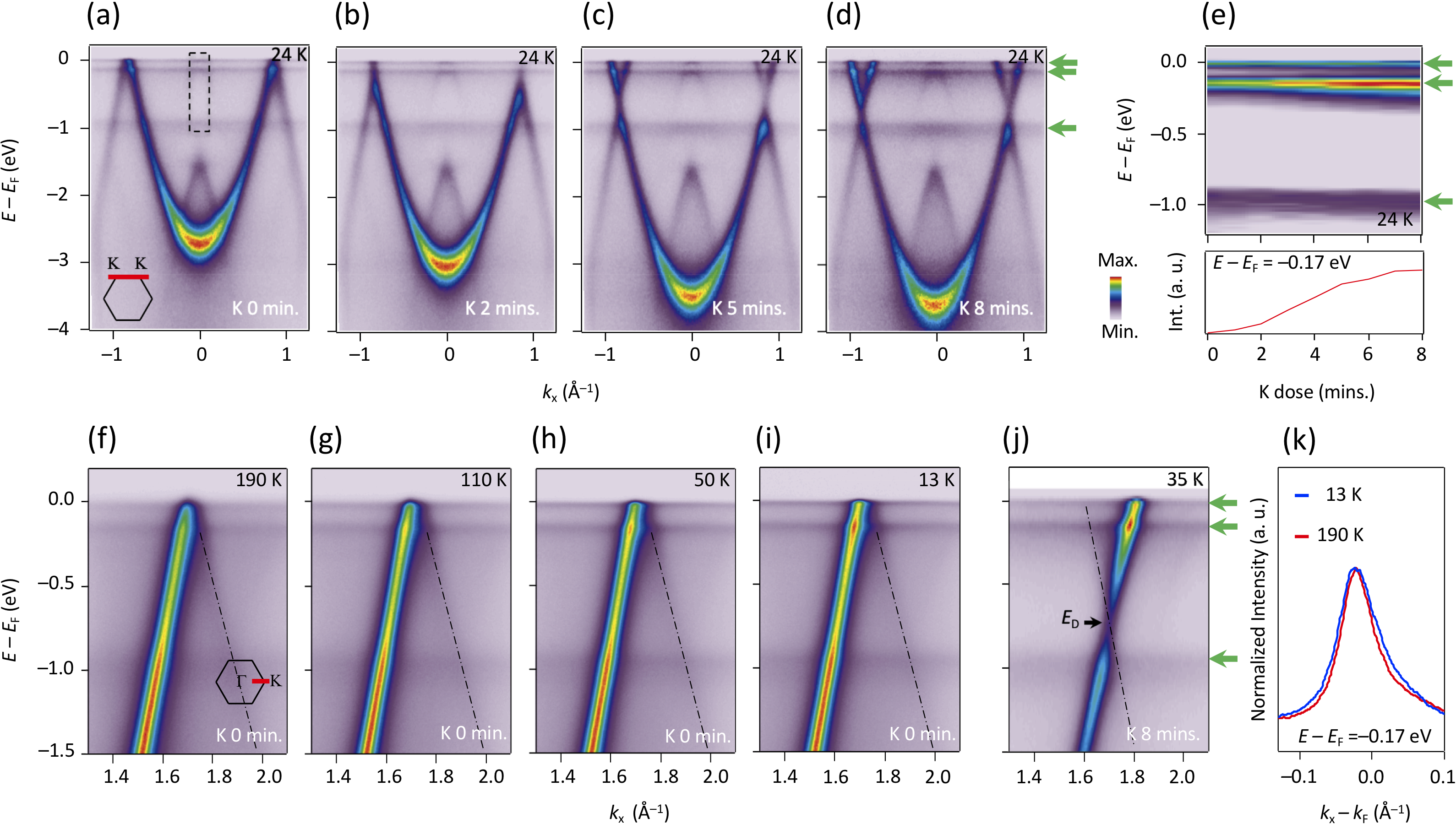}
  \end{center}
  \caption{(a-d) ARPES intensity maps for potassium-deposited graphene on SmB$_6$ measured at 22~K along the K-M-K direction of the graphene unit cell, as denoted in the inset. The green arrows denote the Kondo resonance states. (e) ARPES intensity as a function of potassium dose taken from the area denoted by a dashed rectangle in panel (a). The lower panel shows spectral intensity taken at $E-E_{\rm F}=-0.17~{\rm eV}$ from the ARPES intensity map in the upper panel. (f-i) ARPES intensity maps measured at several different temperatures along the $\Gamma$-K direction of the graphene unit cell, as denoted in the inset. The dash-single dotted line denotes another branch of the conical dispersion of graphene whose intensity is depressed due to the photoemission selection rule~\cite{CK}. (j) An ARPES intensity map measured at 35 K along the $\Gamma$-K direction of the graphene unit cell after 8~minutes of potassium dose on graphene. The green and black arrows denote the Kondo resonance states and the Dirac energy of graphene, respectively. (k) The momentum distribution curves taken at $E-E_{\rm F}=-0.17~{\rm eV}$ for the same pristine graphene at 13~K (blue) and 190~K (red), for comparison.}
  \label{Fig3}
  \end{figure*}

The graphene $\pi$ band also reflects the coupling between graphene and SmB$_6$. Figures~3(f-i) show ARPES intensity maps taken along the $\Gamma$-K direction of the graphene unit cell at different temperatures. The photoelectron intensity of one out of two branches of the conical dispersion is significantly suppressed along this direction~\cite{CK}. Three non-dispersive states at 0.02~eV, 0.17~eV, and 0.96~eV below $E_{\rm F}$ denoted by green arrows are the Kondo resonance states, originating from SmB$_6$. As temperature decreases, the Kondo resonance states increase in spectral intensity, which is also observed in bare SmB$_6$~\cite{Jonathan2}. At the same time, non-trivial features appear in the graphene $\pi$ band. Its spectral intensity at the crossing points with the Kondo resonance states is gradually enhanced upon cooling, that is also observed in electron-doped graphene as shown in Fig.~3(j). The momentum distribution curve (MDC) taken at $E-E_{\rm F}=-0.17~{\rm eV}$, corresponding to the $^{6}H_{7/2}$ final-state multiple, of the 13~K data (Fig. 3(i)) is wider than that of 190~K (Fig. 3(f)), when both of them are normalized for comparison as shown in Fig. 3(k). This is opposite from what can be expected from thermal broadening. These unusual spectral features of the graphene $\pi$ band observed at the crossing points with the Kondo resonance states suggest that they do not remain intact, but are intertwined with each other, e.\,g.\,, to form a hybridized state, that becomes prominent at higher charge carrier density and at lower temperature. 

The hybridization is often manifested as a kink at the crossing point between two states~\cite{Yb,Pt}. Indeed, the graphene $\pi$ band shows interesting dependence on both temperature and charge carrier density in this sense. Figure~4(a) is the energy-momentum dispersions of the graphene $\pi$ band obtained by a Lorentzian fit to the ARPES data taken at several temperatures. The slope of the dispersion over an energy range of $-0.25~{\rm eV} \leq E-E_{\rm F} \leq -0.10~{\rm eV}$ gradually changes upon cooling. Here the dotted line is the energy corresponding to the $^{6}H_{7/2}$ final-state multiplet, that is extracted from the peak position of the energy distribution curve shown on the right hand side obtained by integrating the ARPES intensity map in a range of 2.0~\AA$^{-1} \leq k_{\rm x} \leq 2.1$~\AA$^{-1}$ in Fig.~3(i). More specifically, the 190~K dispersion (the red curve) is almost linear with minute deviation from the linearity that is the characteristics of the graphene $\pi$ band on a metallic substrate~\cite{CuSiegel}. Here, the dashed line is an arbitrary straight line for a guide to the eyes. The dispersion exhibits gradually increasing deviation from the linearity with its maximum at the lowest temperature, e.\,g.\,, 13~K (the purple curve), so that the kink in the graphene $\pi$ band becomes notable at the crossing point with the Kondo resonance state.

  \begin{figure*}[t]
  \begin{center}
  \includegraphics[width=1\columnwidth]{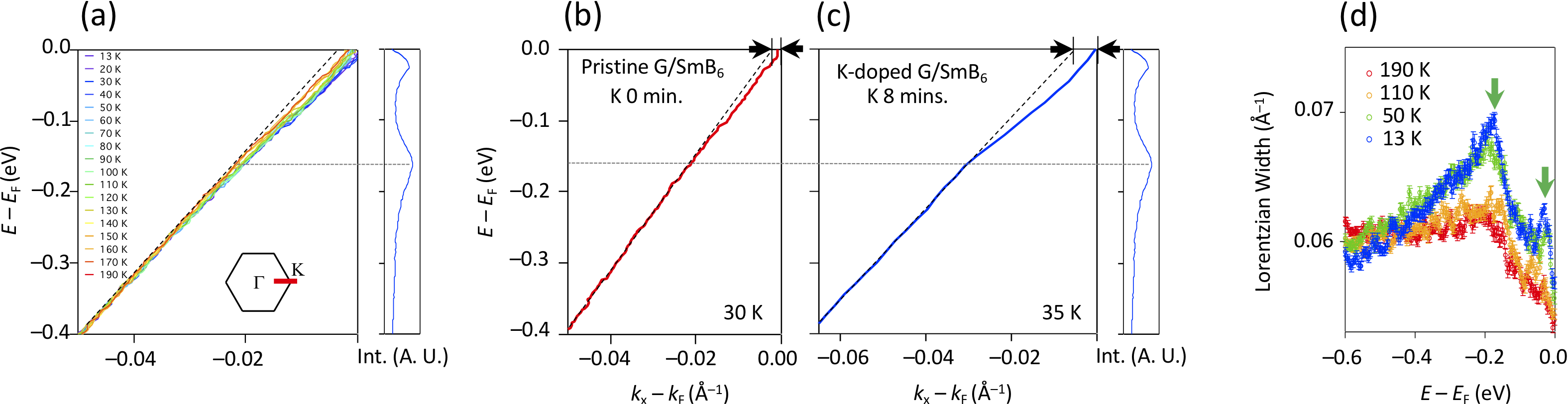}
  \end{center}
  \caption{(a) Energy-momentum dispersions of graphene on SmB$_6$ obtained by a Lorentzian fit to ARPES data taken along the $\Gamma$-K direction of the graphene unit cell at several different temperatures. The dotted line is a guide to the eyes obtained from the peak position of one of the Kondo resonance states. (b-c) An energy-momentum dispersion of graphene on SmB$_6$ at 190~K (red) and 13~K (blue). Right panel is the angle-integrated spectral intensity in a range of 2.0~\AA$^{-1} \leq k_{\rm x} \leq 2.1$~\AA$^{-1}$ in Fig.~3(i). The dashed line is a guide to the eyes to show the deviation from the linearity that is denoted by black arrows. (d) An energy-momentum dispersion of graphene on SmB$_6$ with 8~minutes of potassium dose obtained by a Lorentzian fit to the ARPES data shown in Fig.~3(j). (e) Lorentzian peak width for the ARPES data shown in Figs.~3(f-i). Green arrows denote the energies of the two lowest energy Kondo resonance states. }
  \label{Fig4}
  \end{figure*}

The observed kink is also influenced by the charge carrier density of graphene. Figures~4(b-c) compare the energy-momentum dispersions for pristine and potassium-dosed graphene on SmB$_6$ taken at similar temperature, 30~K and 35~K, respectively. With 8~minutes of potassium dose, the charge carrier density of graphene changes from $n_{\rm h}=2\times10^{11}~{\rm cm}^{-2}$ of holes to $n_{\rm e}=4\times10^{13}~{\rm cm}^{-2}$ of electrons. With the increase of charge carrier density, the strength of the kink is strongly enhanced. Electron-phonon coupling and a band bending effect~\cite{Calandra} might be a possible explanation of the existence of the kink and its enhanced strength upon increasing charge carrier density. However, the observed temperature dependence of the energy-momentum dispersion and the peak width displayed in Fig.~4(d) are not understood within both scenarios, especially when the peak width lacks the characteristic feature of the electron-phonon coupling, e.\,g.\,, step-function-like enhancement by the enhanced phonon emission beyond the phonon energy~\cite{Yb}. Instead, with decreasing temperature, the spectral width of the graphene $\pi$ band shows sharp enhancement at the crossing point with the Kondo resonance states as denoted by green arrows, consistent with the hybridization picture. While the enhanced width of the graphene $\pi$ band around the Kondo resonance states at lower temperature is in contrast to the thermal broadening picture, the width away from the Kondo resonance states, e.\,g.\,, $E-E_{\rm F}=-0.6~{\rm eV}$, shows the thermal broadening effect, i.\,e.\,, increasing width with increasing temperature.

The dependence of the Kondo resonance state and its hybridization with the graphene $\pi$ band on both charge carrier density of graphene and temperature is consistent with the Kondo physics that predicts a stronger effect at lower temperature with higher charge carrier density, providing an experimental evidence of Kondo effect in graphene. Here, graphene plays an important role. The chemical inertness of graphene allows not only graphene itself, but also SmB$_6$ to preserve their intact surface structures, while the thermal decomposition method makes it possible to prepare their interface with perfectly aligned azimuthal orientation. The formation of such a clean and well-defined interface introduces close-packed lattice-type local magnetic impurities from the surface of SmB$_6$ to graphene, providing two important ingredients of the Kondo effect, e.\,g.\,, local magnetic moment from SmB$_6$ and metallic background from graphene, that is predicted to show strong temperature dependence. In addition, the capability to manipulate the charge carrier density of graphene makes it possible to tune the strength of the Kondo effect without modifying the physical properties of both graphene and SmB$_6$. These advantages reveal a clear signature of the 2D lattice-type Kondo effect in graphene in the temperature range where Kondo resonance is observed in SmB$_6$. This is particularly interesting because graphene, unless heavily charge doped~\cite{HwangCe}, is expected to show extremely low $T_{\rm K}$~\cite{YJiang}, suggesting that the Kondo physics of a 3D Kondo insulator is transferred to a 2D non-magnetic Dirac semimetal, that is tuned by temperature and charge carrier density.

In summary, we have demonstrated that an ideal interfacial structure between graphene and a substate of interests can be realized using thermal decomposition of fullerene molecules. The formation of the interface between graphene and SmB$_6$ allows the Kondo effect of the substrate to be transferred to non-magnetic graphene. We also provide an efficient way to tune the interfacial Kondo effect via temperature and charge carrier density of graphene. This approach can be applied to other 3D substrates with exotic properties to realize and manipulate a new type of emergent fermionic quasiparticles that do not exist when graphene stands alone. For example, despite graphene consists of a light element, carbon, strong electronic correlations can be realized in graphene when it is placed on heavy fermionic or multiferroic materials. This will shed light on the fundamental comprehension of correlated phases that appear in the two-dimensional sea of Dirac fermions. Interestingly, SmB$_6$ is also a potential candidate for this purpose. The coupling between almost massless Dirac fermions from graphene and recently reported topologically protected heavy Dirac fermions from the surface of SmB$_6$~\cite{Jonathan2,Jiang,Seunghun} can provide a unique playground to explore a strongly correlated quantum spin Hall phase~\cite{Andreas}.

\section{Experiments}

The surface of a single-crystal SmB$_6$(110) sample was polished and sputter-annealed at 1300~$^{\circ}$C in an ultra-high vacuum prep chamber with a base pressure of 8$\times$10$^{-11}$ Torr. The surface quality was examined by the 1$\times$1 LEED pattern and the electron band structure, corresponding to the bulk-terminated non-polar SmB$_6$(110) surface. The non-polar surface was chosen to focus on the proximity-induced effect, but not on the topological nature of the substrate, in which the polar nature of the surface is supposed to be involved. Fullerene molecules were sublimed out of a crucible at about 300~$^{\circ}$C to deposit on to the thus prepared SmB$_6$ surface. The sample was kept at room temperature during the deposition process. The sample was then flashed to 1300~$^{\circ}$C, which graphitizes the fullerene molecules, leaving only the single-layer graphene as evidenced by its characteristic electron band structure via ARPES. Graphene is estimated to cover most of the SmB$_6$ surface via ARPES {\it xy}-scan and LEED, whose lateral resolution is 50 $\mu$m and 1 mm, respectively. Once the formation of the highly ordered graphene was confirmed, subsequent anneals at lower temperatures, e.\,g.\,, 680-930~$^{\circ}$C, have been employed to remove foreign adsorbates. The flashing and annealing process and the presence of graphene do not affect the key properties of SmB$_6$, as the electron band structure as well as the Kondo resonance of SmB$_6$ persist throughout these processes. Potassium atoms were deposited using a commercial SAES disperser at 30~K in an ultra-high vacuum main chamber with a based pressure of 3$\times$10$^{-11}$ Torr. Both sample preparation and ARPES measurements were performed at the MERLIN beamline 4.0.3 of the Advanced Light Source. Photon energy was tuned from 68~to 86~eV. Sample temperature was controlled from 15~K to 190~K. At 15~K using 68~eV photons, the energy resolution was set to be 7~meV.




\begin{acknowledgement}

This work was supported by the National Research Foundation of Korea (NRF) grant funded by the Korea government (MSIP) (No.~2018R1A2B6004538 and 2020K1A3A7A09080369). The Advanced Light Source is supported by the Office of Basic Energy Sciences of the U.S. Department of Energy under Contract No.~DE-AC02-05CH11231. S.L., H.R., and J.–H.P. acknowledge support by Max Planck POSTECH/Korea Research Initiative, Study for Nano Scale Optomaterials and Complex Phase Materials (No.~2016K1A4A4A01922028) through NRF funded by MSIP of Korea. H.R. also acknowledges support from the KIST Institutional Program (2E29410). C.H. is greatly indebted to J. Kim for her support that has made this study possible. 

\end{acknowledgement}

%
%
%


\end{document}